\documentclass[conference]{IEEEtran}

\usepackage{cite}
\usepackage{amsmath,amssymb,amsfonts}
\usepackage{algorithmic}
\usepackage{graphicx}
\usepackage{textcomp}
\usepackage{xcolor}
\def\BibTeX{{\rm B\kern-.05em{\sc i\kern-.025em b}\kern-.08em
    T\kern-.1667em\lower.7ex\hbox{E}\kern-.125emX}}
\begin{document}

\title{QAOA-Based Pilot Assignment for Cell-Free Massive MIMO Systems
}

\author{
\IEEEauthorblockN{1\textsuperscript{st} Xiaoyu Ma}
\IEEEauthorblockA{
\textit{Department of Electrical and Computer} \\
\textit{Engineering} \\
\textit{Western University} \\
London, ON N6A 5B9, Canada \\
xma535@uwo.ca
}
\and
\IEEEauthorblockN{2\textsuperscript{nd} Fang Fang}
\IEEEauthorblockA{
\textit{Department of Electrical and Computer} \\
\textit{Engineering} \\
\textit{Department of Computer Science} \\
\textit{Western University} \\
London, ON N6A 5B9, Canada \\
fang.fang@uwo.ca
}
\and
\IEEEauthorblockN{3\textsuperscript{rd} Xianbin Wang}
\IEEEauthorblockA{
\textit{Department of Electrical and Computer} \\
\textit{Engineering} \\
\textit{Western University} \\
London, ON N6A 5B9, Canada \\
xianbin.wang@uwo.ca
}
}

\maketitle

\begin{abstract}
Pilot contamination is a major challenge in cell-free massive multiple-input multiple-output (MIMO) systems, where the limited number of orthogonal pilots makes pilot reuse unavoidable. Since the pilot assignment solution space grows exponentially with the number of users, finding the global optimum becomes computationally challenging on classical computers. Recent advances in quantum computing provide a promising approach for solving such large-scale combinatorial optimization problems. In this paper, we reformulate the assignment problem as a quantum-compatible optimization problem, enabling it to be directly solved by the Quantum Approximate Optimization Algorithm (QAOA). Specifically, the pilot contamination objective and assignment constraints are incorporated into the QAOA formulation. This allows the quantum search to focus on valid pilot assignments with low contamination cost. Simulation results show that the proposed method achieves performance close to exhaustive search, demonstrating the potential of quantum-assisted optimization for pilot assignment in future  wireless systems.
\end{abstract}


\section{Introduction}
Cell-free massive multiple-input multiple-output (MIMO) has emerged as a promising architecture for future wireless communication systems, where a large number of distributed access points (APs) jointly serve user equipments (UEs) under the coordination of a central processing unit (CPU)~\cite{7827017, demir2021foundations}. However, accurate channel estimation remains essential for cell-free massive MIMO. In time-division duplexing (TDD) systems, uplink pilots are used for channel estimation, but the limited coherence block length restricts the number of orthogonal pilots. When the number of UEs exceeds the pilot length, pilot reuse becomes unavoidable and leads to pilot contamination. Pilot assignment is an effective approach to mitigating this problem~\cite{9178782, 11160717}. In cell-free massive MIMO systems, the impact of pilot reuse depends on the large-scale fading relationships between UEs and APs. If two UEs are strongly observed by similar APs, assigning them the same pilot may cause severe contamination. Therefore, pilot assignment can be formulated as an optimization problem, where the pilot-sharing cost is determined by pairwise UE interference. For $K$ UEs and $\tau_p$ available pilots, the search space contains $\tau_p^K$ possible assignment patterns. Since the pilot choices of different UEs are coupled through pairwise interference relationships, finding the globally optimal pilot assignment becomes computationally challenging on classical computers as the network size increases.

Recent advances in quantum computing provide a new perspective for solving large-scale combinatorial optimization problems~\cite{preskill2018quantum}. Among different quantum algorithms, the quantum approximate optimization algorithm (QAOA) has attracted much attention due to its hybrid quantum-classical structure and its applicability to optimization problems~\cite{farhi2014quantum, hadfield2019quantum, ma2026eqe}. The basic principle of QAOA is to encode the objective function into a cost Hamiltonian, whose low-energy states correspond to high-quality candidate solutions. Then, a mixer Hamiltonian is applied to drive the quantum state to explore the solution space. Motivated by this mechanism, QAOA provides a potential framework for solving large-scale wireless optimization problems when quantum computing devices become more practical~\cite{9803257, chen2026recursive}.

In this paper, we propose a constraint-aware QAOA-based pilot assignment framework for cell-free massive MIMO systems, enabling the pilot contamination minimization problem to be addressed through quantum-assisted optimization. The main contributions are summarized as follows: (i) we develop a QAOA-compatible formulation for weighted pilot assignment in cell-free massive MIMO systems; (ii) we construct a cost Hamiltonian and a constraint-preserving mixer to encode the pilot contamination objective while maintaining feasible pilot assignments; and (iii) we demonstrate the effectiveness of the proposed framework through simulations, showing pilot assignment performance close to exhaustive search.

\section{System Model and Problem Formulation}
We consider an uplink cell-free massive MIMO system consisting of $M$ distributed APs and $K$ single-antenna UEs. The APs are connected to a CPU, which coordinates pilot assignment and signal processing. The system operates in TDD mode, where uplink pilot signals are first transmitted for channel estimation, followed by uplink data transmission. Since pilot assignment is based on large-scale fading, it is updated only when the user set or large-scale channel conditions change, rather than in every coherence block.

\begin{figure}[htbp]
\centering
\includegraphics[width=1.0\columnwidth]{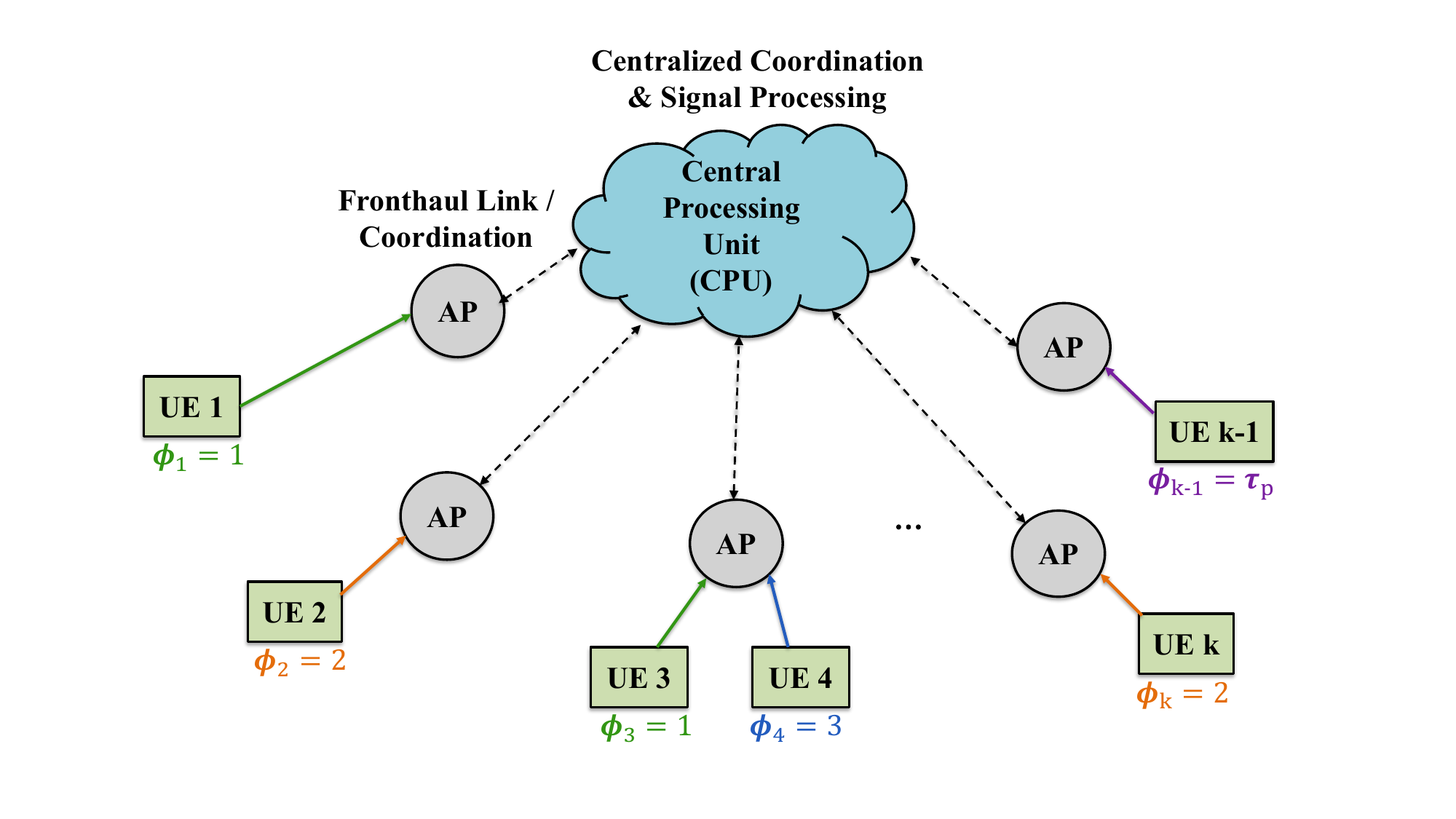}
\caption{Uplink pilot assignment in a cell-free massive MIMO system with centralized coordination.}
\label{system_model}
\end{figure}

Within each coherence block, \(\tau_p\) symbols are used for pilot transmission, providing \(\tau_p\) orthogonal pilots. When \(K>\tau_p\), multiple UEs must reuse the same pilots, which may cause pilot contamination during channel estimation. Therefore, pilot assignment should be designed according to the interference relationships among UEs. Let $\phi_k \in \{1,2,\ldots,\tau_p\}$ denote the pilot assigned to UE $k$, where $k=1,\ldots,K$. The complete pilot assignment vector is given by $\boldsymbol{\phi}
=[\phi_1,\phi_2,\ldots,\phi_K]$. For any pair of users $j$ and $k$, we define the pilot reuse indicator as
\begin{equation}
    I_{j,k} = \mathbb{I}\{\phi_j=\phi_k\}.
\end{equation}
An example of uplink pilot assignment and pilot reuse in a cell-free massive MIMO system is illustrated in Fig.~\ref{system_model}.

The cost of pilot reuse depends on how strongly two users interfere with each other during channel estimation. In a cell-free massive MIMO system, this interference is related to whether the two users are strongly observed by the same APs. Let $\beta_{m,k}$ denote the large-scale fading coefficient between AP $m$ and UE $k$. The large-scale fading vector of UE $k$ is defined as $\boldsymbol{\beta}_k
= [\beta_{1,k},\beta_{2,k},\ldots,\beta_{M,k}]^T$. To quantify the potential pilot contamination between users $j$ and $k$, we define the user interference weight as
\begin{equation}
    \rho_{j,k} = \sum_{m=1}^{M} \beta_{m,j}\beta_{m,k}.
    \label{interference_weight}
\end{equation}
A larger $\rho_{j,k}$ means that users $j$ and $k$ are strongly received by similar APs. Therefore, if they reuse the same pilot, they are more likely to cause severe pilot contamination. In contrast, a smaller $\rho_{j,k}$ indicates weaker channel overlap, meaning that reusing the same pilot would have a lower interference impact for these two users.

The objective of pilot assignment is to allocate one pilot to each user such that strongly interfering users are less likely to share the same pilot. If two users $j$ and $k$ are assigned the same pilot, this pilot-sharing pair contributes an interference cost proportional to their interference weight $\rho_{j,k}$. Therefore, the total pilot contamination cost is defined as
\begin{equation}
    C(\boldsymbol{\phi}) = \sum_{1\leq j<k\leq K} \rho_{j,k} \mathbb{I}\{\phi_j=\phi_k\}.
\end{equation}
The weighted pilot assignment problem is formulated as
\begin{equation}
    \begin{aligned}
\min_{\boldsymbol{\phi}} \quad
& C(\boldsymbol{\phi})
=
\sum_{1\leq j<k\leq K}
\rho_{j,k}
\mathbb{I}\{\phi_j=\phi_k\} \\
\text{s.t.} \quad
& \phi_k \in \{1,2,\ldots,\tau_p\}, \quad k=1,\ldots,K.
\end{aligned}
\end{equation}
The objective assigns a pairwise cost whenever two users reuse the same pilot, and the cost is determined by their interference weight $\rho_{j,k}$. Therefore, minimizing $C(\boldsymbol{\phi})$ reduces the total interference weight among user pairs sharing pilots.

The total number of possible pilot assignment patterns is $\tau_p^K$, which grows exponentially with the number of users. Moreover, the pilot choice of each user is coupled with the choices of other users through the pairwise interference weights. Therefore, the weighted pilot assignment problem is a coupled combinatorial optimization problem and becomes increasingly challenging as the network size grows.

\section{QAOA-Based Pilot Assignment}

To solve the weighted pilot assignment problem, we formulate it within a QAOA-based optimization framework. The pilot-sharing cost is encoded into a cost Hamiltonian, while a feasibility-preserving mixer Hamiltonian is designed to explore only valid pilot assignments. After parameter optimization, the quantum state is measured and decoded into the final pilot assignment.

\subsection{Quantum Encoding and Hamiltonian Construction}

The pilot assignment variable $\phi_k \in \{1,2,\ldots,\tau_p\}$ is first transformed into a binary representation. For each UE $k$ and each pilot index $q$, we define a binary variable
\begin{equation}
    x_{k,q} = 
    \begin{cases}
    1, & \text{if UE } k \text{ is assigned pilot } q,\\
    0, & \text{otherwise}.
    \end{cases}
\end{equation}
Since each UE must be assigned exactly one pilot, the binary variables satisfy the one-hot constraint
\begin{equation}
    \sum_{q=1}^{\tau_p} x_{k,q}=1,\quad k=1,\ldots,K.
\end{equation}
Under this encoding, the pilot reuse indicator between users $j$ and $k$ can be written as
\begin{equation}
    \mathbb{I}\{\phi_j=\phi_k\} = \sum_{q=1}^{\tau_p} x_{j,q}x_{k,q}.
\end{equation}
This equality holds because users $j$ and $k$ share the same pilot if and only if there exists a pilot $q$ such that both $x_{j,q}=1$ and $x_{k,q}=1$. Therefore, the weighted pilot assignment cost becomes
\begin{equation}
    C(\mathbf{x}) = \sum_{1\leq j<k\leq K} \rho_{j,k} \sum_{q=1}^{\tau_p} x_{j,q}x_{k,q}.
\end{equation}
This formulation converts the original pilot assignment problem into a binary quadratic optimization problem over the one-hot feasible space.

To map this cost function to a quantum Hamiltonian, each binary variable $x_{k,q}$ is represented by one qubit. Let $Z_{k,q}$ denote the Pauli-$Z$ operator acting on the qubit corresponding to $x_{k,q}$. The Pauli-Z operator is diagonal in the computational basis and distinguishes the two basis states $|0\rangle$ and $|1\rangle$ through their eigenvalues. In our encoding, this allows the qubit state $|0\rangle$ to represent $x_{k,q}=0$ and the qubit state $|1\rangle$ to represent $x_{k,q}=1$. The binary variable is therefore represented by the operator
\begin{equation}
\hat{x}_{k,q} = \frac{I-Z_{k,q}}{2}.
\end{equation}
Based on the above binary representation, the QAOA cost Hamiltonian is given by
\begin{equation}
    H_C = \sum_{1\leq j<k\leq K} \rho_{j,k} \sum_{q=1}^{\tau_p} \hat{x}_{j,q}\hat{x}_{k,q}.
\end{equation}
For any feasible pilot assignment state $|\mathbf{x}\rangle$, the cost Hamiltonian directly encodes the weighted pilot contamination cost, i.e., $H_C|\mathbf{x}\rangle=C(\mathbf{x})|\mathbf{x}\rangle$. Since $H_C$ is diagonal in the computational basis, different feasible pilot assignments are assigned different energy values according to their pilot contamination costs. Therefore, minimizing the expectation value of $H_C$ is equivalent to searching for a pilot assignment with a low total pilot contamination cost.

However, the one-hot constraint must also be preserved during the QAOA evolution. A standard $X$-mixer flips individual qubits independently and may generate infeasible states, where one UE is assigned no pilot or multiple pilots. To avoid this issue, we design a feasibility-preserving mixer that only transfers the selected pilot of each UE from one pilot index to another. For UE $k$, we define the mixer Hamiltonian as
\begin{equation}
    H_{M,k} = \sum_{1\leq q<r\leq \tau_p} \frac{1}{2} \left(X_{k,q}X_{k,r} + Y_{k,q}Y_{k,r} \right),
\end{equation}
where $X_{k,q}$ and $Y_{k,q}$ are Pauli-$X$ and Pauli-$Y$ operators acting on the qubit associated with $x_{k,q}$. The overall mixer Hamiltonian is given by
\begin{equation}
    H_M = \sum_{k=1}^{K} H_{M,k}.
\end{equation}
The term $\frac{1}{2}(X_{k,q}X_{k,r}+Y_{k,q}Y_{k,r})$ exchanges the excitation between the two pilot qubits $q$ and $r$ of the same UE. Therefore, it changes the pilot selected by UE $k$ while preserving the number of selected pilots for that UE.

\subsection{QAOA Evolution and Solution Decoding}

Given the $H_C$ and $H_M$, QAOA constructs a parameterized quantum state by alternating between the corresponding cost and mixer evolutions. The cost unitary is defined as
\begin{equation}
    U_C(\gamma) = e^{-i\gamma H_C},
\end{equation}
where $\gamma$ is a variational parameter. This operator introduces phase shifts according to the pilot contamination cost of each assignment. The mixer unitary is defined as
\begin{equation}
    U_M(\beta) = e^{-i\beta H_M},
\end{equation}
where $\beta$ is another variational parameter. Since $H_M$ preserves the one-hot constraint, the mixer evolution explores only valid pilot assignment states.

The initial state is chosen as a uniform superposition over all feasible pilot assignments. For each UE $k$, we define
\begin{equation}
    |s_k\rangle = \frac{1}{\sqrt{\tau_p}} \sum_{q=1}^{\tau_p} |q\rangle_k,
\end{equation}
where $|q\rangle_k$ denotes the one-hot state in which UE $k$ is assigned pilot $q$. The overall initial state is
\begin{equation}
    |\psi_0\rangle = \bigotimes_{k=1}^{K} |s_k\rangle.
\end{equation}
After $p$ QAOA layers, the parameterized quantum state is given by
\begin{equation}
    |\psi_p(\boldsymbol{\gamma},\boldsymbol{\beta})\rangle = \prod_{\ell=1}^{p} U_M(\beta_\ell) U_C(\gamma_\ell) |\psi_0\rangle,
\end{equation}
where $\boldsymbol{\gamma}=[\gamma_1,\ldots,\gamma_p]$ and $\boldsymbol{\beta}=[\beta_1,\ldots,\beta_p]$ are variational parameters. These parameters are optimized by minimizing the expected value of the cost Hamiltonian:
\begin{equation}
    \min_{\boldsymbol{\gamma},\boldsymbol{\beta}} \langle \psi_p(\boldsymbol{\gamma},\boldsymbol{\beta})| H_C |\psi_p(\boldsymbol{\gamma},\boldsymbol{\beta})\rangle.
\end{equation}
The optimization follows a hybrid quantum-classical procedure. For a given parameter set, the quantum circuit prepares the QAOA state and estimates the expected cost from measurement outcomes. A classical optimizer then updates the parameters to reduce the expected cost.

After parameter optimization, the optimized QAOA state is measured in the computational basis to generate a set of candidate pilot assignments. Each measurement outcome corresponds to a one-hot binary assignment $\mathbf{x}$, which is decoded into a pilot assignment vector $\boldsymbol{\phi}$. Specifically, UE $k$ is assigned pilot $q$ if $x_{k,q}=1$. The cost of each sampled assignment is evaluated, and the measured assignment with the lowest cost is selected as the final pilot assignment solution.

\section{Simulation Results}

We evaluate the proposed QAOA-based pilot assignment method using Qiskit's noiseless statevector simulator. The simulated cell-free massive MIMO system consists of $M=50$ APs uniformly distributed over a $1\,\mathrm{km}\times1\,\mathrm{km}$ area, with large-scale fading generated by a distance-dependent path-loss model with exponent $3.7$. The interference weights are calculated according to \eqref{interference_weight}. We set $\tau_p=3$ and consider $K\in\{4,5,6,7\}$, with each result averaged over 30 independent network realizations. The QAOA parameters are optimized using COBYLA with five random initializations and at most 100 iterations, followed by 2048 samples from the optimized state. Due to the computational cost of classical statevector simulation, the current evaluation is limited to \(K\leq 7\), while NISQ noise and hardware validation are left for future work.

\begin{figure}[htbp]
  \centering
  \includegraphics[width=1.0\linewidth]{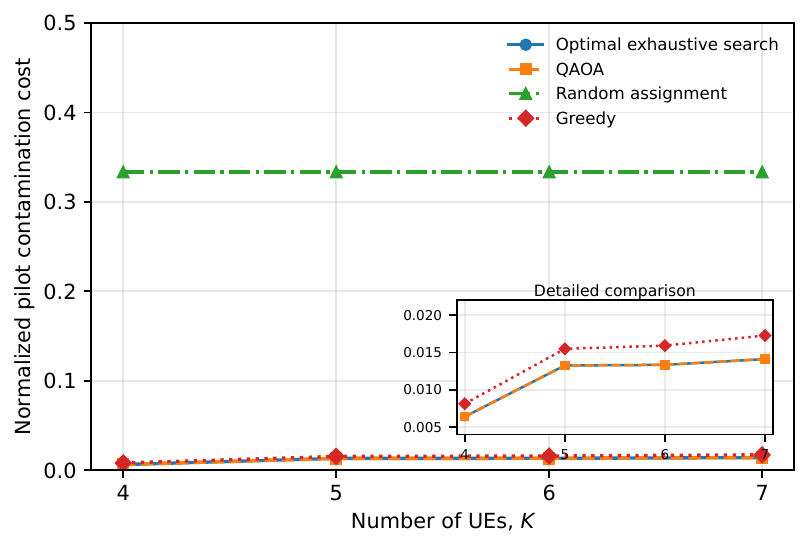}
  \caption{Normalized pilot contamination cost versus the number of UEs with $p=3$.}
  \label{cost_k}
\end{figure}

We first evaluate the pilot-assignment quality as the number of UEs increases. The QAOA depth is fixed at $p=3$, and the proposed method is compared with exhaustive search, random assignment, and greedy assignment. The pilot contamination cost is normalized as $\bar{C}(\boldsymbol{\phi})=C(\boldsymbol{\phi})/\sum_{1\leq j<k\leq K}\rho_{j,k}$. For QAOA, the lowest-cost feasible assignment among the sampled solutions is selected as the final output. As shown in Fig.~\ref{cost_k}, random assignment gives the highest normalized contamination cost, while QAOA and greedy assignment significantly reduce the cost. The inset shows that QAOA closely approaches the exhaustive-search optimum for all considered values of $K$, while slightly outperforming the greedy baseline. Although the improvement over greedy is modest at the evaluated scale, the results demonstrate that QAOA can obtain higher-quality pilot assignments than a strong classical heuristic.

\begin{figure}[htbp]
  \centering
  \includegraphics[width=1.0\linewidth]{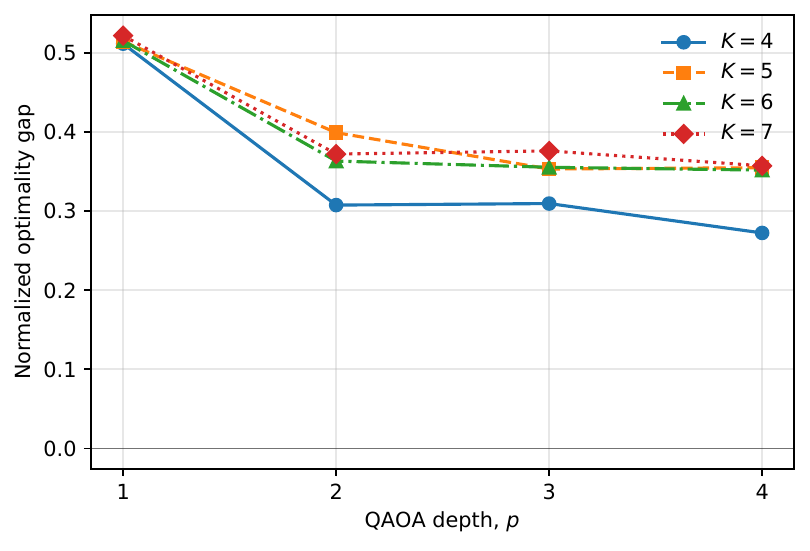}
  \caption{Normalized optimality gap versus QAOA depth for different numbers of UEs.}
  \label{gap_depth}
\end{figure}

We further evaluate the effect of the QAOA circuit depth using the normalized optimality gap, defined as $G_p=(\mathbb{E}[\bar{C}_p]-\bar{C}^{\star})/(\bar{C}_{\mathrm{rand}}-\bar{C}^{\star})$, where $\mathbb{E}[\bar{C}_p]$ is the optimized expected normalized QAOA cost, $\bar{C}^{\star}$ is the exhaustive-search optimum, and $\bar{C}_{\mathrm{rand}}$ is the average random-assignment cost. As shown in Fig.~\ref{gap_depth}, increasing the depth from $p=1$ to $p=2$ substantially reduces the gap for all considered values of $K$, showing that additional QAOA layers improve the representation of low-cost assignments. Further increasing the depth provides smaller gains, and the curves gradually saturate. For larger problem sizes, especially $K=7$, the remaining gap is higher, suggesting that larger pilot-assignment problems may require deeper circuits or more effective parameter optimization.

\section{Conclusion}
In this paper, we construct a constraint-aware QAOA formulation for pilot assignment in cell-free massive MIMO systems. The weighted pilot-sharing objective is encoded into the cost Hamiltonian, while the feasible assignment structure is preserved by the mixer Hamiltonian for the one-hot constraint. This formulation provides a QAOA-compatible representation of pilot contamination minimization. It also indicates the potential of constraint-aware quantum optimization for wireless resource allocation problems. Future work will focus on improving the scalability of the proposed formulation and evaluating its performance under more practical quantum computing settings.


\begin{thebibliography}{00}

\bibitem{7827017}
H. Q. Ngo, A. Ashikhmin, H. Yang, E. G. Larsson, and T. L. Marzetta,
``Cell-free massive MIMO versus small cells,''
IEEE Trans. Wireless Commun., vol. 16, no. 3, pp. 1834--1850,
Mar. 2017.

\bibitem{demir2021foundations}
{\"O}. T. Demir, E. Bj{\"o}rnson, and L. Sanguinetti,
``Foundations of user-centric cell-free massive MIMO,''
Found. Trends Signal Process., vol. 14, nos. 3--4,
pp. 162--472, 2021.

\bibitem{9178782}
S. Buzzi, C. D'Andrea, M. Fresia, Y.-P. Zhang, and S. Feng,
``Pilot assignment in cell-free massive MIMO based on the Hungarian algorithm,''
IEEE Wireless Commun. Lett., vol. 10, no. 1, pp. 34--37,
Jan. 2021.

\bibitem{11160717}
D. Pereira-Ruis{\'a}nchez, {\'O}. Fresnedo, D. P{\'e}rez-Ad{\'a}n,
and L. Castedo,
``C-Footprints: A statistic-based clustering for pilot allocation in
cell-free massive MIMO,''
in Proc. IEEE Int. Conf. Commun. (ICC), 2025, pp. 2303--2308.

\bibitem{preskill2018quantum}
J. Preskill,
``Quantum computing in the NISQ era and beyond,''
Quantum, vol. 2, Art. no. 79, Aug. 2018.

\bibitem{farhi2014quantum}
E. Farhi, J. Goldstone, and S. Gutmann,
``A quantum approximate optimization algorithm,''
2014, arXiv:1411.4028. [Online]. Available:
https://arxiv.org/abs/1411.4028

\bibitem{hadfield2019quantum}
S. Hadfield, Z. Wang, B. O'Gorman, E. G. Rieffel, D. Venturelli,
and R. Biswas,
``From the quantum approximate optimization algorithm to a quantum
alternating operator ansatz,''
Algorithms, vol. 12, no. 2, Art. no. 34, Feb. 2019.

\bibitem{ma2026eqe}
X. Ma, F. Fang, X. Xie, X. Wang, and L. Hanzo,
``EQE-QAOA: An equivalence-preserving qubit efficient framework for
combinatorial optimization,''
2026, arXiv:2604.18285. [Online]. Available:
https://arxiv.org/abs/2604.18285

\bibitem{9803257}
J. Cui, Y. Xiong, S. X. Ng, and L. Hanzo,
``Quantum approximate optimization algorithm based maximum likelihood
detection,''
IEEE Trans. Commun., vol. 70, no. 8, pp. 5386--5400,
Aug. 2022.

\bibitem{chen2026recursive}
K.-C. Chen, H. Matsuyama, W.-H. Huang, and Y. Yamashiro,
``Recursive QAOA for interference-aware resource allocation in wireless
networks,''
2026, arXiv:2602.07483. [Online]. Available:
https://arxiv.org/abs/2602.07483

\end{thebibliography}
\end{document}